\documentclass[12pt,preprint]{aastex}

\begin{document}

\newcommand\etal{et al. }

\def\etal{et al.\ }
\def\msun{M_{\sun}}
\def\rsun{R_{\sun}}
\def\lsun{L_{\sun}}
\def\kms{\rm \, km \, s^{-1}}
\def\ha{H$\alpha \;$}
\def\ecs{\rm erg \, cm^{-2} \, s^{-1}}
\def\micron{$\mu$m}

\title{The 24-Micron View of Embedded Star Formation in NGC 7129}
\author{J. Muzerolle \altaffilmark{1}, S. T. Megeath \altaffilmark{2},
R. A. Gutermuth \altaffilmark{3}, L. E. Allen \altaffilmark{2},
J. L. Pipher \altaffilmark{3}, L. Hartmann \altaffilmark{2},
K. D. Gordon \altaffilmark{1}, D. L. Padgett \altaffilmark{5},
A. Noriega-Crespo \altaffilmark{5},
P. C. Myers \altaffilmark{2}, G. G. Fazio \altaffilmark{2},
G. H. Rieke \altaffilmark{1}, E. T. Young \altaffilmark{1},
J. E. Morrison \altaffilmark{1},
D. C. Hines \altaffilmark{1,4}, K. Y. L. Su \altaffilmark{1},
C. W. Engelbracht \altaffilmark{1}, \& K. A. Misselt \altaffilmark{1}}
\altaffiltext{1}{Steward Observatory, University of Arizona, 933 N. Cherry
Ave., Tucson, AZ 85721 (jamesm@as.arizona.edu)}
\altaffiltext{2}{Harvard-Smithsonian Center for Astrophysics, Mail Stop 42,
60 Garden Street, Cambridge, MA 02138}
\altaffiltext{3}{Department of Physics and Astronomy, University of Rochester,
Rochester, NY 14627}
\altaffiltext{4}{Space Science Institute, 4750 Walnut Street, Suite 205,
Boulder, CO 80301}
\altaffiltext{5}{{\it{Spitzer}} Science Center, Caltech, Pasadena, CA 91125}

\begin{abstract}

We present observations of the star formation region NGC 7129 taken
with the Multiband Imaging Photometer for {\it{Spitzer}} (MIPS).
A significant population of sources, likely pre-main sequence
members of the young stellar cluster, is revealed outside
the central photoionization region.  Combining with Infrared Array
Camera (IRAC) and ground-based near-infrared images, we have obtained
colors and spectral energy distributions for some 60 objects.
The [3.6]-[4.5] vs. [8]-[24] color-color plane shows sources clustered
at several different loci, which roughly correspond to the
archetypal evolutionary sequence Class 0, I, II, and III.  We obtain
preliminary classifications for 36 objects,
and find significant numbers of both Class I and II objects.
Most of the pre-main sequence candidates are
associated with the densest part of the molecular cloud surrounding
the photoionization region, indicating active star
formation over a broad area outside the central cluster.
We discuss three Class II candidates that exhibit
evidence of inner disk clearing, which would be some of the youngest
known examples of a transition from accretion to optically thin
quiescent disks.

\end{abstract}


\section{Introduction}

A complete census of young stellar objects in a wide variety of
star forming regions is important to advancing our understanding
of star and planet formation processes.  The prevailing paradigm
places young stellar objects into an evolutionary sequence spanning
the first few million years of their existence (Adams et al. 1987):
the youngest embedded protostars surrounded by infalling envelopes
and growing accretion disks (Class 0/I objects); pre-main sequence
(PMS) stars with less active accretion disks (Class II); 
PMS stars with no disks or optically thin remnant dust (Class III).
The characteristics of this evolutionary sequence
inform the origins of the initial mass function and the formation
and frequency of planetary systems.
By studying a large sample of different stellar nurseries,
we can begin to examine in detail the effects of the star forming
environment on these processes.

Mid-infrared observations are essential for identifying and characterizing
excess dust emission from circumstellar envelopes and disks.
Previous surveys such as IRAS have been able to study in detail only
the nearest star forming regions such as Taurus (e.g. Kenyon et al. 1990;
Kenyon \& Hartmann 1995) and $\rho$ Oph (e.g. Wilking et al. 1989;
Luhman \& Rieke 1999), at distances of 140 and 160 pc, respectively.
The {\it{Spitzer Space Telescope}}, with the combination
of unprecedented sensitivity and spatial resolution provided by
the Infrared Array Camera (IRAC) and Multiband Imaging Photometer
for {\it{Spitzer}} (MIPS), has the capability to characterize regions
as distant as $\sim 1$ kpc, significantly increasing the statistics
of PMS object properties as a function of stellar
environment, mass, and age.

The 24 {\micron} channel of MIPS in particular provides a robust diagnostic
of the presence of circumstellar envelopes around Class 0/I objects,
whose relatively cold dust emission is characterized by a rising spectral
energy distribution (SED) through the mid-infrared.  It is also important for
probing emission from accretion disks around Class II objects, being most
sensitive to structure at $\sim 1-5$ AU (the region of gas giant formation
in the solar nebula), such as
changes in the disk surface height due to dust settling from
grain growth (Miyake \& Nakagawa 1995).
Observations at 24 {\micron} are also essential for identifying
inner gaps in disks, as indicated by the lack
of dust emission at shorter wavelengths.  Such objects are indicative
of a transition between the Class II and III stages that may be
brought about by grain growth/coagulation in the inner disk.
Finally, the 24{\micron} channel provides the primary
means of identifying optically thin dust disks that may be
the remnants of earlier accretion activity, or
(especially at older ages) ``debris" generated by planetesimal collisions.

NGC 7129 offers a prime example of a very young embedded region
($\sim 1$ Myr), with significant molecular material undergoing
active star formation.  The cloud has already been partially
disrupted by winds and photoionization from newly-formed massive
stars (Miskolczi et al. 2001), creating a reflection nebula visible
in the optical.  A significant population
of pre-main sequence objects within the reflection nebula
has been indicated by ground-based near-infrared observations
(Hodapp 1994); however, a complete census of
young objects at all PMS evolutionary phases is lacking.
We present MIPS observations of this cluster, focusing on the
24 {\micron} imaging to identify protostellar envelopes and accretion disks,
and obtain a preliminary picture of the PMS population outside
the central reflection nebula.

\section{Observations}

\begin{figure}
\plotone{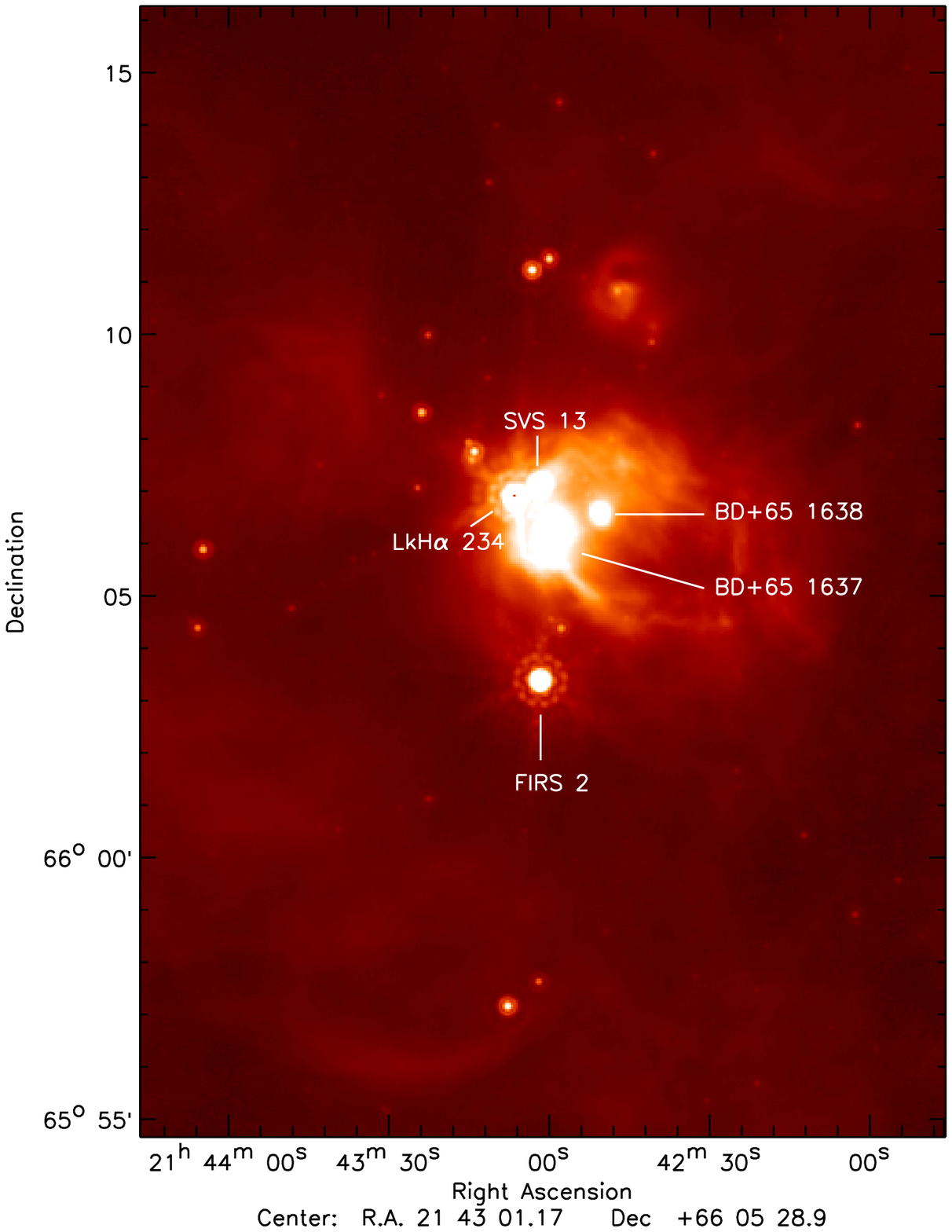}
\caption{A portion of the 24$\mu$m scan map of NGC 7129.
\label{map}}
\end{figure}

The region was mapped with MIPS using scan mode (Rieke et al. 2004),
covering a total area of 15' by 30' common to all three detector arrays.
The map consists of 6 scan legs with half-array cross-scan offsets,
providing the necessary redundancy so that all points on
the sky in the region of interest were placed on side ``A"
of the 70 {\micron} array (Rieke et al.).  Medium scan rate
was employed, providing a total effective exposure
time per pixel of 80 seconds at 24 {\micron}, 40 seconds at 70
{\micron}, and 8 seconds at 160 {\micron}.  The data were reduced and
mosaicked using the instrument team in-house Data Analysis Tool,
which includes per-exposure calibration of all detector
transient effects, dark current subtraction, and
flat fielding/illumination correction (Gordon et al. 2004).
Coaddition and mosaicking of individual frames included
applying distortion corrections and cosmic ray rejection.
A portion of the final 24 {\micron} mosaic is shown in Figure~\ref{map}.

\section{Results}

The 24 {\micron} image of NGC 7129 shows bright extended emission
in the center that corresponds to the reflection nebula
seen in the optical.  The primary sources of photoionization,
the B3 stars BD+65 1637 and BD+65 1638, are not detected directly,
but extended halos of warm dust are seen at their positions.
Known PMS stars LkH$\alpha$ 234 and SVS 13,
to the northeast of BD+65 1637, are completely saturated
in the PSF core.  The previously identified far-infrared source
FIRS 2 appears to the south of BD+65 1637;
it is slightly saturated in the central pixel.  The cluster of
pre-main sequence stars located within the reflection nebula
(Gutermuth et al. 2004) is largely unseen at 24 {\micron}
because of the bright background emission and lower spatial resolution
of the instrument.  However, a fairly large number of point sources
is detected outside the central region, primarily in an arc to the north,
east, and south.  Since MIPS does not have
the sensitivity to detect most stellar photospheres at
the distance of NGC 7129 (which we take to be 1 kpc; Racine 1968),
most of these sources likely exhibit infrared excesses,
which we explore in more detail below.

We measured photometry on point sources using daophot with PSF-fitting
(the flux for FIRS 2 was recovered by fitting the PSF wings;
the other two saturated sources were unrecoverable).
A total of 182 point sources was measured over the entire map
at 24 {\micron}, four at 70 {\micron}, and none at 160 {\micron}.
The latter two arrays suffer from the very strong extended emission
and saturate over large areas, which significantly curtailed
the sensitivity to weaker point sources.  We thus focus
our analysis on the 24 {\micron}
data.  Typical errors, dominated by uncertainties
in the absolute calibration, are within $\sim 10\%$
at 24 {\micron} and $\sim 20\%$ at 70 {\micron}.
Finally, fluxes were converted into magnitudes referenced to
the Vega-system using a zero-point (7.3 Jy) derived from
the Vega spectrum.

\begin{figure}
\epsscale{0.9}
\plotone{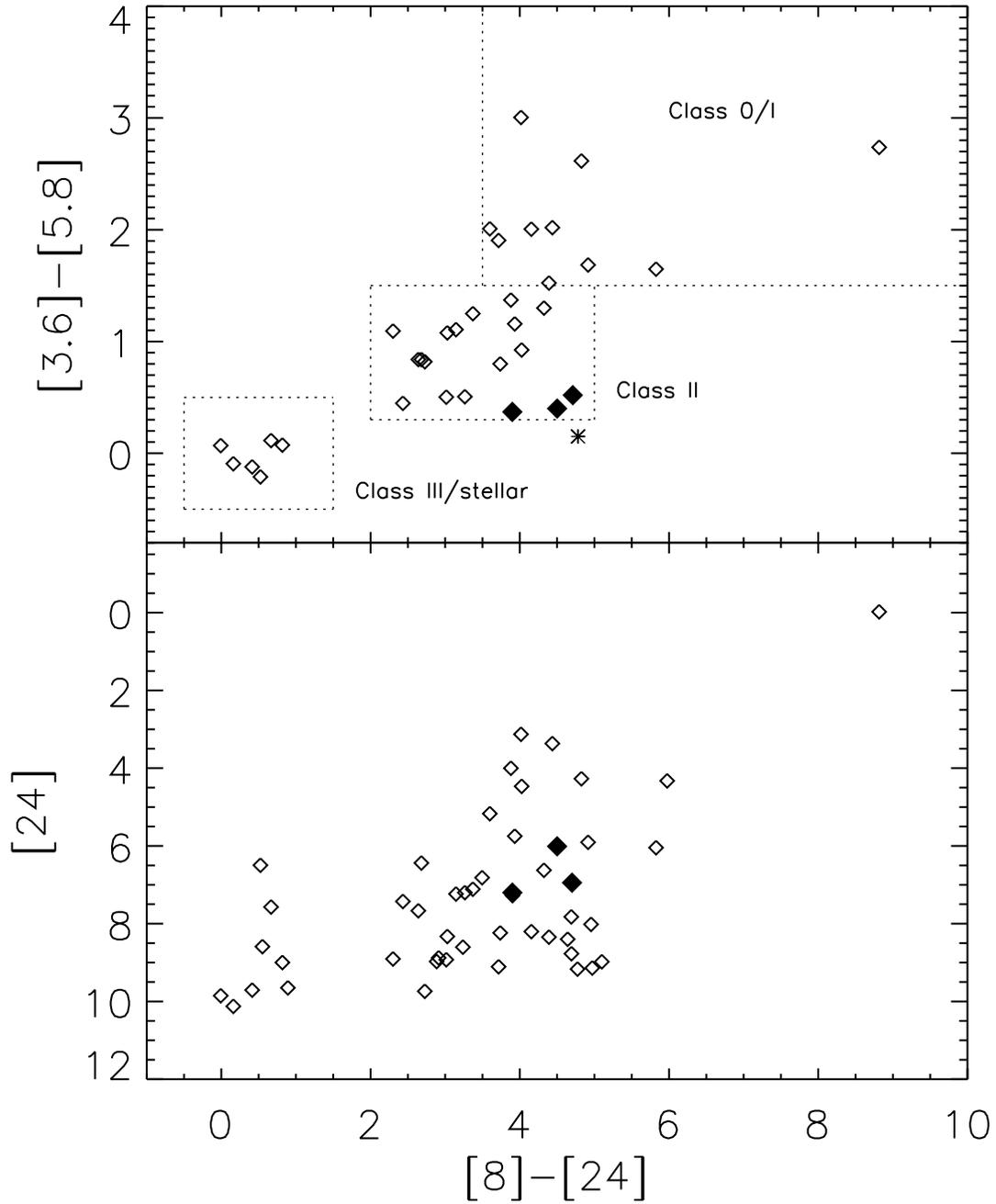}
\caption{Color-color and color-magnitude diagrams for all sources
detected in all appropriate bands.  Approximate regions for
basic object classifications are delineated in the upper panel.
The asterisk represents the T Tauri star TW Hya, as determined
from IRAC and IRAS 25 {\micron} fluxes; solid diamonds are objects
with similar SEDs, and may be transition objects with evidence
for a gap in the inner disk (see discussion).
\label{diagrams}}
\end{figure}

We have combined our MIPS measurements with ground-based
$JHK$ and IRAC 4-channel photometry (see Gutermuth et al. 2004
for details).  The observations are not simultaneous,
so we are potentially subject to uncertainties from variability;
however, emission at the longer wavelengths originates mostly
from a larger area of circumstellar material farther from
the central star, so we do not expect significant variability
at 24 {\micron}.  With these data, we can determine infrared colors
and construct spectral energy distributions for our sources,
and begin to probe the characteristics of the young stellar
objects in NGC 7129.  Along these lines, we present the ([8]-[24])
vs. ([3.6]-[5.8]) color-color and ([8]-[24]) vs. [24]
color-magnitude diagrams in Figure~\ref{diagrams} for all sources
detected in these bands.
The ([8]-[24]) color in particular is very sensitive to excesses,
since photospheric colors should be close to zero for all spectral
types.

Figure~\ref{diagrams} shows that almost all the sources
plotted have a significant infrared excess.
A small group of 6 sources is clustered around
([3.5]-[5.8]) $\sim 0$, $0 \lesssim
([8]-[24]) \lesssim 1$, and is probably a mixture of pure photospheres
and perhaps weak 24 {\micron} excesses.  The remainder
of the sources have strong ([8]-[24]) excesses and moderate-to-strong
([3.5]-[5.8]) excesses; these are likely to be Class I objects
with envelopes or Class II objects with optically thick disks.
The source with the most extreme colors corresponds to FIRS 2,
an outflow source and probably a Class 0 protostar given its
cold, massive envelope as seen at $mm$ wavelengths (Eiroa et al. 1998)
(however, we caution that much of the flux in the IRAC bands may
be contaminated by shocked emission from the outflow).
There is an interesting gap at ([8]-[24]) $\sim 1-2.5$,
which suggests a lack of objects having remnant disks
with inner regions that are no longer optically thick in the infrared.

A more detailed look at the spectral energy distributions is
required to better disentangle the object classifications.
In Figure~\ref{seds}, we show typical SEDs spanning the range
of characteristics of our total sample; coordinates and magnitudes for
these sources are listed in Table 1.  Class I objects
are characterized by a rising SED through near- and mid-infrared
wavelengths, typically peaking at 60-100 {\micron} (Lada 1987).
The top panel of Figure~\ref{seds} shows 4 such objects
(including 70 {\micron} measurements, at which wavelength
the SEDs are still rising).  The SED of a known Class I object
in Taurus, L1551-IRS5, is shown for comparison.  The lower panel
shows examples of candidate Class II sources (including two with
evidence for a large inner disk hole - see discussion).
The Taurus Class II median SED (D'Alessio et al. 1999)
is shown for comparison.  By examining the SEDs, we have determined
approximate regions in the color-color plane corresponding
to each of the object classifications, as shown in Figure~\ref{diagrams}.

The similarity of the SEDs of our new
candidates to known PMS objects strongly suggests a PMS origin
for most if not all of them.  Spectra are needed
to definitively weed out non-PMS objects that can also have infrared
excesses, such as galaxies or planetary nebulae.
Given the significant extinction of the
molecular cloud and the relatively small solid angle subtended
by the map, we do not expect such contamination to be significant.

\section{Discussion}

\subsection{The Pre-Main Sequence Population}

We have created a preliminary inventory of young stellar objects
in NGC 7129 using our infrared SEDs.  The PMS candidates
discussed here comprise only the population located outside of
the central reflection nebula, since most of the stars there are
hidden by the associated strong extended emission at 24 {\micron}.  
To avoid ambiguity, we have restricted classification to sources
detected at 24 {\micron} and at least three of the IRAC bands.
A total of 39 objects meet this criterion, out of 49 sources
located within the overlapping field of view of the IRAC observations
(all of which are detected in at least one IRAC band).
The remaining 24 {\micron} sources are worthy of
follow-up study.  In particular, 7 objects are not detected at
5.8 and 8 {\micron}; some of these may be edge-on disk sources,
which are predicted to have double-humped SEDs with a large dip
between $\sim 3 - 20$ {\micron} (D'Alessio et al. 1999).

Of the 39 sources detected in at least four bands, we find that
about a third (13) are candidate protostars with circumstellar
envelopes.  One of these, the source FIRS 2, is probably a Class 0 object,
as mentioned above.  The other 12, with less extreme colors, are likely
Class I objects.  Included in this category are two objects
with relatively flat spectral slopes (often known as ``flat-spectrum"
sources), which may be at an evolutionary stage intermediate
between Class I and II.
Another 18 objects appear to be Class II
sources; most if not all of these are probably classical T Tauri stars,
given their relatively low infrared luminosities.
Six objects have SEDs that are closer to pure photospheres
(as also indicated in the color-color diagram),
and may be a combination of Class III objects, background giants,
or foreground dwarfs.  A few have slightly larger (8-24) colors
which may indicate the presence of an optically thin remnant disk.
Spectral types are needed to disentangle this group of objects.
Finally, three sources have more complicated SEDs that we do not
classify; two of them are very faint and appear to the southwest of
the reflection nebula, where the extinction is much smaller,
and hence may be galaxies.

At first glance, our preliminary results indicate a large
fraction of Class I to Class II sources in the outer region
of NGC 7129.  However, we are missing the core cluster members
located inside the reflection nebula.  The statistics are
also affected to an uncertain degree by completeness at low luminosities.
Objects in both classes span a fairly large range
of luminosity, but we are likely missing some number of very low-luminosity
objects, as well as sources with extreme extinction.
The weakest Class II sources we detect are $\sim 10$ times
less luminous than the scaled Taurus median SED that corresponds
to a median source luminosity $L \sim 0.8 \; L_{\odot}$ (see Fig.~\ref{seds}).
Similarly low-luminosity protostars do not appear in our classified
subsample, probably because their fluxes at the shorter IRAC wavelengths
are below our detection threshold.  If we add in
the pre-main sequence candidates found from the IRAC and near-infrared
data alone (Allen et al. 2004; Gutermuth et al. 2004; Megeath et al. 2004),
which includes both low-luminosity sources and the central
cluster members, we end up with a total of about 20 Class 0/I
and 80 Class II objects, resulting in a fraction very similar
to that seen in Taurus (Kenyon \& Hartmann 1995).

\begin{figure}
\plotone{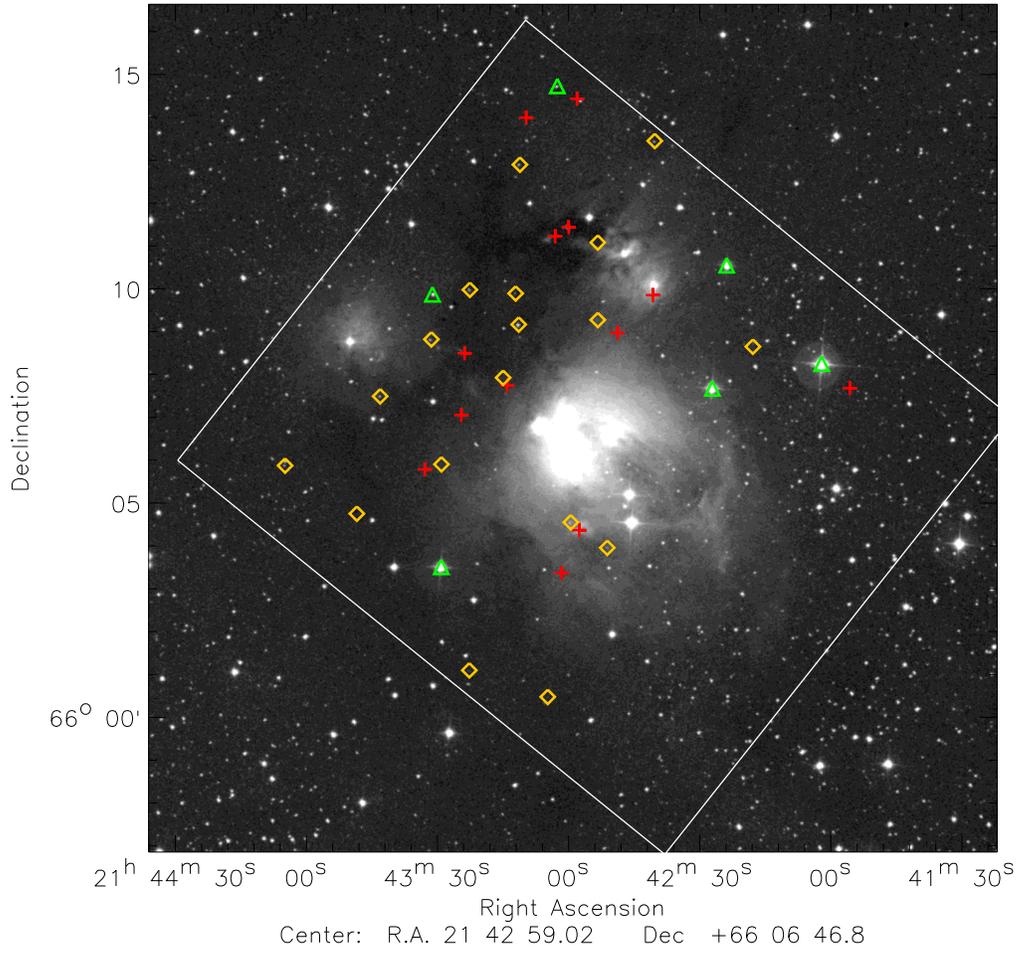}
\caption{Optical DSS image of NGC 7129 overlaid with symbols showing
the positions of candidate young stellar objects.  Red pluses: Class 0/I
objects; yellow diamonds: Class II objects; green triangles:
Class III/dwarfs/giants.  The overlapping IRAC field of view
is shown with a white box.
\label{positions}}
\end{figure}

Another interesting result is the spatial distribution of
our subsample of 36 classified objects, shown in Figure~\ref{positions}.
As is seen for the sample as a whole, most of these objects are
located in an arc sweeping north-east-south of the central
reflection nebula.  This area largely coincides
with the densest parts of the molecular cloud, as seen in the
optical image and in the gas distribution traced by $^{13}$CO and
C$^{18}$O emission (Miskolczi et al. 2001; Ridge et al. 2003).
Based on the presence of protostars in this region, we argue that
active star formation must be dispersed
over a fairly large area ($\sim 3$ pc), and not concentrated
in the center as is usually observed.  The winds and photoionizing
flux of the central B-stars have cleared out most of the molecular
material from the central region, preventing formation of new protostars,
and may eventually contribute to the cessation of continuing
star formation in the surrounding cloud within the next few Myr.

\subsection{Transition Objects: Inner Disk Gaps?}

\begin{figure}
\epsscale{0.75}
\plotone{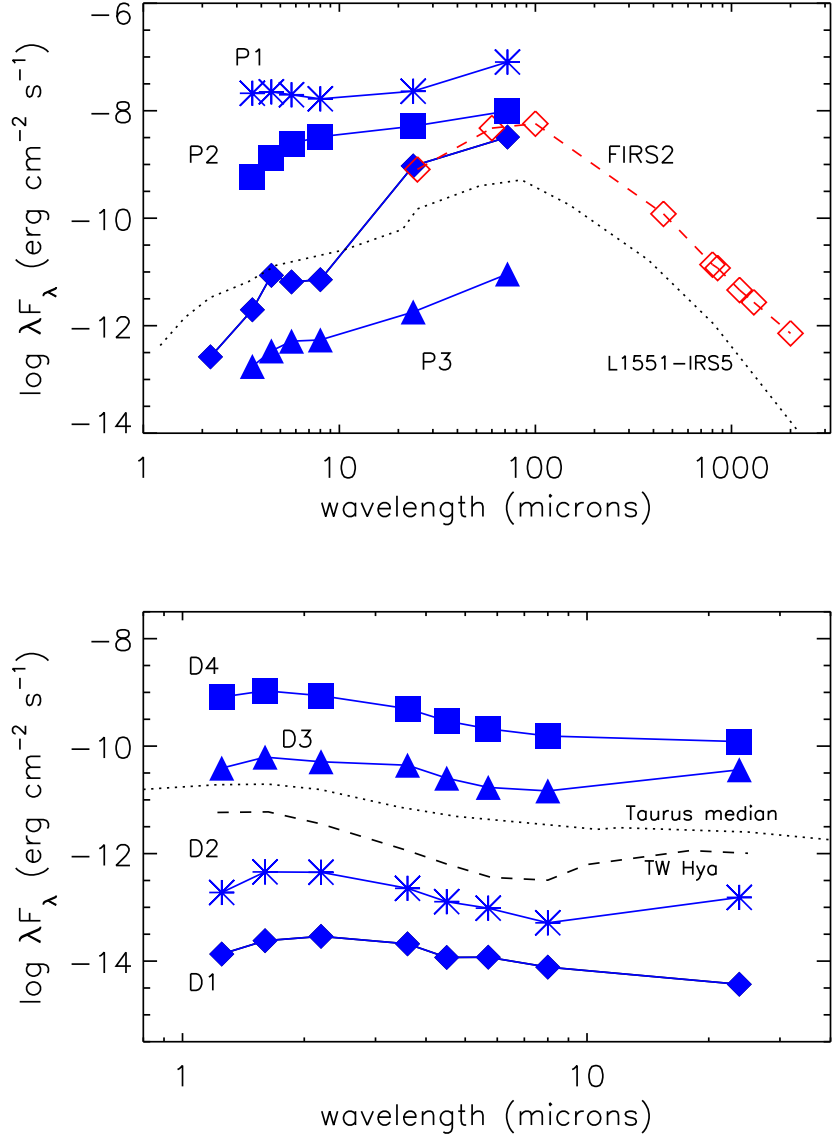}
\caption{Spectral energy distributions for selected sources.
({\it{Top panel}})  Protostellar candidates: shown are the 4 sources
detected at both 24 and 70 {\micron}.  Solid symbols connected by solid lines
are IRAC/MIPS fluxes; open symbols connected by a dashed line are
IRAS (color corrected assuming a $T=50$ K blackbody spectrum) and
ground-based sub-mm/mm observations of FIRS 2 (Eiroa et al. 1998;
Font et al. 2001).
The following sources are shifted along the y-axis for clarity:
P1 (+3 dex); P2 (+2 dex); P3 (-1 dex).  The SED of Taurus Class I
object L1551-IRS5, scaled to 1 kpc, is shown with
the dotted line (Osorio et al. 2003, and references therein).
Note that the fluxes for FIRS 2 at $K$ through 5.8 {\micron} are likely
contaminated by shocked emission from outflows (see Gutermuth et al. 2004).
({\it{Bottom panel}}) Class II candidates.  Source D1 has been shifted
by -1.5 dex, D2 by -1 dex, D3 by +2 dex, and D4 by +1 dex
along the y-axis.  The median SED for Taurus class II objects
(D'Alessio et al. 1999) and the SED for TW Hya
(data from Webb et al. 1999, Jayawardhana et al. 1999,
and IRAS and IRAC observations),
both scaled to 1 kpc, are shown with the dotted and dashed lines,
respectively.
\label{seds}}
\end{figure}

We detect three objects whose SEDs show little or no
excess at wavelengths $\lesssim 8$ {\micron}, but strong excesses
at 24 {\micron} (two of these are shown in Fig.~\ref{seds}, objects D2 and D3).
Such SEDs are strongly suggestive
of disks with inner gaps, such as the 10 Myr-old Class II object
TW Hya (Calvet et al. 2002; also shown in Fig.~\ref{seds}).
One of the central goals of circumstellar disk studies is
the determination of the timescales for disk evolution,
which can constrain the mechanism(s) for planet
formation.  Evidence so far indicates that primordial disks
do not last longer than $\sim 10$ Myr; however, a large fraction of
PMS stars do not show evidence for disks even at ages of 1-3 Myr
(Haisch et al. 2001; Gutermuth et al. 2004).
Thus, the timescale of the transition from optically thick
primordial disk to no disk or regenerative ``debris" disk
is thought to be fairly short.  Objects which show disks with
an inner gap may be in this transition phase, with clearing
of disk material related to planetesimal accretion.
Unfortunately, only a very few such objects
have been identified, most of which are older than $\sim 1$ Myr.
Our candidates may provide important new
examples of this transition at a very young age.

\acknowledgements

This work is based in part on observations made with the {\it{Spitzer Space
Telescope}}, which is operated by the Jet Propulsion Laboratory,
California Institute of Technology under NASA contract 1407.
Support for this work was provided by NASA through Contract Number
960785 issued by JPL/Caltech.

\clearpage

\begin{deluxetable}{lllccccccccc}
\tabletypesize{\small}
\tablecaption{Selected MIPS sources}
\tablewidth{0pt}
\tablehead{
\colhead{Object} & \colhead{$\alpha$(J2000)} & \colhead{$\delta$(J2000)} &
\colhead{J} & \colhead{H} & \colhead{K} & \colhead{3.6} & \colhead{4.5} &
\colhead{5.8} & \colhead{8.0} & \colhead{24} & \colhead{70}}
\startdata
FIRS 2 & 21 43 01.66 & 66 03 24.0 & \nodata & \nodata & 16.35 & 12.67 & 10.35 & 9.93 & 8.79 & -0.02 & -5.01 \\
P1 & 21 43 00.00 & 66 11 28.3 & \nodata & \nodata & \nodata & 10.10 & 9.32 & 8.73 & 7.88 & 4.00 & -1.00 \\
P2 & 21 43 03.12 & 66 11 15.4 & \nodata & \nodata & \nodata & 11.50 & 9.88 & 8.49 & 7.14 & 3.13 & -1.22 \\
P3 & 21 43 24.00 & 66 08 31.6 & \nodata & \nodata & \nodata & 12.81 & 11.37 & 10.19 & 9.10 & 4.27 & -1.14 \\
D1 & 21 42 17.52 & 66 08 40.6 & 17.38 & 16.01 & 15.00 & 13.86 & 13.77 & 13.04 & 12.47 & 9.74 & \nodata \\
D2 & 21 42 53.28 & 66 11 06.7 & 15.76 & 14.06 & 13.27 & 12.53 & 12.43 & 12.01 & 11.66 & 6.95 & \nodata \\
D3 & 21 43 31.68 & 66 08 51.0 & 14.17 & 13.13 & 12.55 & 11.67 & 11.53 & 11.16 & 10.46 & 7.20 & \nodata \\
D4 & 21 42 59.51 & 66 04 35.0 & 13.27 & 11.90 & 11.13 & 10.44 & 10.01 & 9.60 & 9.12 & 6.55 & \nodata
\enddata
\end{deluxetable}

\end{document}